\documentclass[runningheads]{llncs}
\usepackage[T1]{fontenc}
\usepackage{graphicx}
\usepackage{booktabs}
\usepackage{subfig}
\usepackage{listings}
\usepackage{xcolor}
\usepackage{hyperref}
\usepackage{tikz}
\usepackage{multirow}
\usepackage{multicol}
\usepackage{amsmath}
\usepackage{array}
\usepackage{makecell}
\usepackage{seqsplit}
\usepackage{url}
\usepackage{soul}
\usepackage{marvosym}

\definecolor{myred}{RGB}{255,0,0}
\definecolor{myblue}{RGB}{0,0,255}

\newcommand*\fullcirc{
\begin{tikzpicture}[baseline=-0.6ex]
    \fill (0,0) circle (0.8ex);
\end{tikzpicture}
}

\newcommand*\emptycirc{
\begin{tikzpicture}[baseline=-0.6ex]
    \draw[black,fill=white] (0,0) circle (0.8ex);
\end{tikzpicture}
}

\newcommand{\circled}[1]{\tikz[baseline=(char.base)]{
            \node[shape=circle,draw,inner sep=0.5pt] (char) {#1};}}

\definecolor{lightgray}{rgb}{0.90, 0.90, 0.90}

\newcommand{\attackvector}[1]{%
  \tikz[baseline=(char.base)]{
    \node[shape=circle, draw=black, fill=lightgray, inner sep=0.1pt, font=\scriptsize] (char) {#1};}}

\newcommand{\attackvectorscript}[1]{%
  \tikz[baseline=(char.base)]{
    \node[shape=circle, draw=black, fill=lightgray, inner sep=0.1pt, font=\scriptsize] (char) {#1};}}

\usepackage{color}

\begin{document}

\title{The Hidden Dangers of Public Serverless Repositories: An Empirical Security Assessment}

\titlerunning{The Hidden Dangers of Public Serverless Repositories}

\author{Eduard Marin\inst{1}\textsuperscript{(\Letter)} \and
Jinwoo Kim\inst{2} \and
Alessio Pavoni\inst{1} \and
Mauro Conti\inst{3,4} \and
Roberto Di Pietro\inst{5}}

\authorrunning{E.\ Marin et al.}
%

\institute{Telefonica Research, Spain \\
\email{\{eduard.marinfabregas, alessio.pavoni\}@telefonica.com}
\and
Kwangwoon University, Republic of Korea \\
\email{jinwookim@kw.ac.kr}
\and
University of Padua, Italy \\
\email{mauro.conti@unipd.it}
\and
 Örebro University, Sweden
\and
King Abdullah University of Science and Technology, Saudi Arabia \\
\email{roberto.dipietro@kaust.edu.sa}
}


%
%




%
\maketitle              

\begin{abstract}


Serverless computing has rapidly emerged as a prominent cloud paradigm, enabling developers to focus solely on application logic without the burden of managing servers or underlying infrastructure. Public serverless repositories have become key to accelerating the development of serverless applications. However, their growing popularity makes them attractive targets for adversaries. Despite this, the security posture of these repositories remains largely unexplored, exposing developers and organizations to potential risks. In this paper, we present the first comprehensive analysis of the security landscape of serverless components hosted in public repositories. We analyse 2,758 serverless components from five widely used public repositories popular among developers and enterprises, and 125,936 Infrastructure as Code (IaC) templates across three widely used IaC frameworks. Our analysis reveals systemic vulnerabilities including outdated software packages, misuse of sensitive parameters, exploitable deployment configurations, susceptibility to typo-squatting attacks and opportunities to embed malicious behaviour within compressed serverless components. Finally, we provide practical recommendations to mitigate these threats.


\end{abstract}

\section{Introduction}

Serverless computing has become a highly compelling cloud paradigm that abstracts infrastructure management tasks (e.g., load balancing and scaling) from tenants, allowing them to focus entirely on application development~\cite{serverless1,serverless2,DBLP:journals/corr/abs-1902-03383,10.1186/s13677-022-00347-w}. In serverless architectures, applications are implemented as a set of small, interdependent \emph{functions}, each designed to perform a specific task. These serverless functions can communicate with one another and integrate with cloud services like event triggers, message queues or object storage to support a wide range of applications. Serverless computing offers automatic scaling in response to workload demands and follows a pure pay-per-use pricing model, where tenants are billed only for the resources consumed during execution. Due to these advantages, major cloud providers, such as AWS~\cite{AWSlambda}, Microsoft~\cite{Azuremicrosoft}, Google~\cite{Googlefunctions}, IBM~\cite{ibm} and Alibaba~\cite{alibaba}, have incorporated serverless computing into their service offerings.

As serverless adoption has grown, numerous \emph{public serverless repositories} have emerged, enabling developers to share serverless components. These repositories host a wide range of serverless components, many of which have been downloaded thousands to millions of times. However, their increasing popularity has also made them attractive targets for adversaries~\cite{prevasio,cryptomining}. A key problem is the lack of transparency surrounding the security practices in these repositories. Most repositories provide little to no information about the security checks performed, the approval policies enforced or how security responsibilities are divided among contributing developers, the users who download serverless components and the repository administrators who publish them. For instance, Red Hat Quay claims “\emph{to continuously scan containers for vulnerabilities}”~\cite{Red}, while AWS states “\emph{all applications published by AWS are reviewed to ensure license compliance and code quality}”~\cite{FAQs}. Although these claims suggest some level of scrutiny, they are highly generic and offer little detail about the scope, rigour or consistency of the security checks applied. Crucially, we argue that such claims may create a false sense of security, leading users to believe that the serverless components they download have been thoroughly vetted and are safe to integrate without further verification, thus increasing the risk of supply chain attacks~\cite{ladisa2023sok}.

To the best of our knowledge, we present the first comprehensive study of the security state surrounding public serverless repositories. We focus on two fundamental yet previously under-explored research questions: (i) Do public serverless repositories introduce application-level security risks? (\textbf{RQ1}); and, (ii) Does the dynamic configuration and deployment model of serverless computing give rise to novel attack vectors? (\textbf{RQ2}). To address {RQ1}, we analyse the prevalence of outdated third-party libraries with known vulnerabilities in serverless components and investigate the potential for embedding malicious behaviour in components distributed as compressed archives. To address {RQ2}, we conduct a detailed analysis of Infrastructure as Code (IaC) templates to identify possible misconfigurations. Additionally, we discover three sensitive parameters in Docker run commands that can be exploited for malicious purposes and evaluate these repositories’ susceptibility to typo-squatting attacks.

To this end, we collect and analyse 2,758 serverless components from five widely-used public repositories: (i) Docker Hub~\cite{dockerhub}; (ii) GitHub~\cite{git}; (iii) AWS Serverless Application Repository (SAR)~\cite{awsrepo}; (iv) Serverless Framework~\cite{Plugins}; and, (v) Red Hat Quay~\cite{Red}. Our selection includes one repository dedicated to \emph{serverless plugins} and four hosting \emph{serverless functions}, spanning both well-maintained platforms (e.g., AWS SAR and Red Hat Quay) and highly popular but less regulated ecosystems such as Docker Hub and GitHub. Additionally, we analyse 125,936 IaC templates from three widely used frameworks: Terraform~\cite{terraform}, AWS CloudFormation~\cite{aws_cloudformation} and AWS Serverless Application Model (SAM)~\cite{aws_sam}.





 \newpage

\noindent\textbf{Contributions.} We summarize our key contributions as follows:

\begin{itemize}
\item We conduct a large-scale security analysis of 2,758 serverless components from major public repositories, including AWS SAR, Docker Hub, GitHub, Red Hat Quay and Serverless Framework, along with 125,936 IaC templates across Terraform, AWS CloudFormation and AWS SAM.

\item We reveal two new attack vectors: the insertion of malicious behaviour into compressed serverless components and the presence of misconfigurations within IaC templates. Additionally, we discover three previously undocumented sensitive parameters in Docker run commands.

\item We provide a set of actionable recommendations aimed at improving the security posture of public serverless repositories and guiding best practices for repository administrators, developers and users.

\end{itemize}

\noindent \textbf{Responsible disclosure.} We notified the maintainers of the affected repositories, providing detailed descriptions of the issues and recommendations for mitigation.

\section{Background and Motivation}\label{sec:background}

\subsection{Serverless Deployment Models}

Serverless functions can be deployed using various methods, including: (i) packaging the function as a Docker container image~\cite{lambdacontainers,googlecloudcontainers,azurecontainers}, (ii) uploading a pre-packaged ZIP folder containing the functions' code and dependencies~\cite{azurezip,googlezip,lambdazip}, (iii) writing code directly in the cloud provider’s console editor~\cite{AWSCLI}, (iv) using YAML templates or configuration files that specify the deployment details, resources and permissions of the serverless functions to be deployed and (v) using Infrastructure as Code (IaC) frameworks, such as AWS CloudFormation, to automate the deployment and management of serverless functions. Regardless of the method, developers remain responsible for providing the application code. In recent years, it has become increasingly common to accelerate development by integrating components from public repositories. However, this introduces significant security risks, as discussed in the next section.

\subsection{Attack Vectors in Public Serverless Repositories}

We identify five primary attack vectors (\attackvector{V1}–\attackvector{V5}) that pose significant security risks to public serverless repositories. It is important to note that some attack vectors are relevant only to specific deployment methods (see Table~\ref{tab:attack_applicability}).

\vspace{0.02in}

\noindent\attackvector{\textbf{V1}} \textbf{Vulnerable third-party libraries.} Serverless components typically rely on third-party libraries, many of which contain known vulnerabilities that introduce security risks~\cite{blackhat}. Even when patches are available, outdated libraries often remain in use for extended periods, offering adversaries ample opportunities to exploit those weaknesses~\cite{ladisa2023sok}. The rapid development cycles inherent to serverless applications make them particularly vulnerable to these risks.

{
\renewcommand{\arraystretch}{1.3}
\begin{table}[t]
\caption{The applicability of attack vectors by deployment model\\(\protect\fullcirc Applicable, \protect\emptycirc Not applicable)}
\centering
\resizebox{\textwidth}{!}{%
\begin{tabular}{l c c c c c}
\toprule
\multirow{2}{*}{\textbf{Attack Vectors}} & \multicolumn{5}{c}{\textbf{Deployment Model}} \\ \cmidrule{2-6} 

& \makecell{Container\\Image} 
& \makecell{Pre-packaged\\Zip} 
& \makecell{Console\\Editor} 
& \makecell{YAML Templates\\Conf. Files} 
& \makecell{IaC\\Frameworks} 
\\ \midrule

\attackvectorscript{V1} Vulnerable third-party libraries 
& \fullcirc & \fullcirc & \fullcirc & \fullcirc & \fullcirc \\

\attackvectorscript{V2} Malicious serverless components
& \fullcirc & \fullcirc & \emptycirc & \emptycirc & \emptycirc \\

\attackvectorscript{V3} Sensitive parameters in Docker run commands
& \fullcirc & \emptycirc & \emptycirc & \emptycirc & \emptycirc \\

\attackvectorscript{V4} Misconfigurations in IaC templates
& \emptycirc & \emptycirc & \emptycirc & \fullcirc & \fullcirc \\

\attackvectorscript{V5} Typo-squatting attacks
& \fullcirc & \fullcirc & \fullcirc & \fullcirc & \fullcirc \\

\bottomrule
\end{tabular}%
}
\label{tab:attack_applicability}
\end{table}
}


\noindent\attackvector{\textbf{V2}} \textbf{Malicious serverless components.} Many repositories allow anyone to upload serverless components after a simple registration process, enabling adversaries to easily distribute malicious components~\cite{franco2023forensic,10.1145/3579856.3590329}. Most serverless platforms also accept pre-packaged compressed archives bundling code and dependencies, which can facilitate more advanced attacks. These formats can further obfuscate malicious behaviour, making detection by traditional security tools significantly more difficult.


\noindent\attackvector{\textbf{V3}} \textbf{Sensitive parameters in Docker run commands.} Some repositories (e.g., Docker Hub) allow contributors to provide execution instructions specifying how their components should be run. Adversaries can exploit this feature and include malicious \texttt{Docker run commands} that contain risky parameters (e.g., \texttt{--privileged} or \texttt{--pid=host}). Users who download these components are likely to follow the provided (malicious) \texttt{Docker run commands}, potentially compromising container isolation and jeopardizing the security of the underlying host.


\noindent\attackvector{\textbf{V4}} \textbf{Misconfigurations in IaC templates.} IaC tools (e.g., AWS CloudFormation) are widely used to specify and deploy serverless functions in production environments. They enable developers to declaratively define the serverless functions and their associated resources, configurations, permissions and policies in a structured and repeatable manner through templates. It is common practice for developers to contribute their IaC templates and reuse those shared by others. However, to date, no systematic study has investigated whether these templates contain misconfigurations or assessed the practical security consequences of such misconfigurations.


\noindent\attackvector{\textbf{V5}} \textbf{Typo-squatting attacks.} Various naming conventions such as Docker’s Fully Qualified Image Identification (FQID)~\cite{277154} and AWS’s Amazon Resource Names (ARNs)~\cite{ARNS} are used to uniquely identify serverless components within repositories. Because developers often enter these names manually (e.g., via terminal or editor), typographical errors are common. Adversaries can exploit this by registering malicious components with names that closely resemble those of popular or trusted ones. This typo-squatting technique leverages human error to surreptitiously distribute malicious serverless components, increasing the likelihood of accidental installation and execution by unsuspecting users.

\subsection{Threat Model}



We consider adversaries capable of uploading vulnerable (\attackvector{V1}) or malicious components (\attackvector{V2}) to these repositories, posing significant risks to unsuspecting users who use them. In doing so, adversaries can also supply execution instructions that include \texttt{Docker run commands} with sensitive parameters (e.g., \texttt{--privileged}) (\attackvector{V3}), exploiting the tendency of many users to follow the provided instructions~\cite{10.1007/978-3-030-58951-6_13}. Similarly, adversaries can upload IaC templates with dangerous configurations to public repositories, causing any developer who uses them to unknowingly misconfigure their serverless applications and inadvertently expose critical information (\attackvector{V4}). Finally, when uploading components to these repositories, adversaries can select component names that closely resemble popular entries in the repository (\attackvector{V5}), aiming to exploit typographical errors made by developers when retrieving serverless components~\cite{277154}.



\section{Security Analysis Framework for Public Serverless Repositories}
\label{methodology}



In this section, we describe the process we used to discover and retrieve serverless components from public repositories, and we evaluate their susceptibility to the five attack vectors considered in this paper.



\vspace{0.02in}

\noindent\textbf{\circled{1} Data collection.} To automate the extraction of serverless component data from the selected repositories, we developed custom web scrapers using frameworks such as BeautifulSoup~\cite{beautifulsoup} and Selenium~\cite{selenium}. Using these scrapers, we collected key metadata, including: (i) the \textit{component name}, (ii) the associated \textit{pull command} or \textit{GitHub URL} and (iii) the recommended \textit{execution instructions} (when available). In some cases, this process required performing authenticated queries and adhering to repository-imposed request limits. For repositories hosting both serverless and non-serverless components, we applied a two-step filtering mechanism to eliminate non-relevant images and keep only the serverless
components. We first configured our crawlers to perform queries using the keyword ‘serverless’ and then examined the metadata associated with each component to detect the presence of a `serverless.yml' file. This methodology was inspired by the approach used by Eskandani et al.~\cite{9463099}.

\vspace{0.02in}

\noindent\textbf{\circled{2} Vulnerability analysis.} Next, we performed a security analysis of the libraries included in the retrieved serverless components using Trivy~\cite{trivy} and Grype~\cite{grype}, two widely used and open-source vulnerability scanners. Both tools extract metadata, package information and libraries from container images or source code, and cross-reference them against multiple vulnerability databases~\cite{trivygrype}. The identified vulnerabilities are classified into five severity levels according to their CVSS 3.1 scores: (i) \emph{Critical} (9.0–10.0), (ii) \emph{High} (7.0–8.9), (iii) \emph{Medium} (4.0–6.9), (iv)~\emph{Low} (0.1–3.9) and (v) \emph{Unknown} (excluded from the analysis). Each serverless component was scanned separately with both tools. Using multiple scanners helps account for tool variability, as each may rely on different vulnerability databases and detection heuristics, potentially identifying distinct sets of vulnerabilities.

\newpage

\noindent\textbf{\circled{3} Hiding malicious behaviour in compressed serverless components.} To investigate whether compression can be exploited to conceal malicious behaviour in serverless components, we utilised VirusTotal~\cite{virustotal}, a widely used platform that aggregates results from numerous antivirus engines. We hypothesized that adversaries could leverage common compression formats to evade detection~\cite{azurezip,googlezip,lambdazip}. To test this, we created serverless components compressed with popular compression formats, including both benign samples and variants embedded with malware. Subsequently, we submitted these samples to VirusTotal and analysed the detection rates reported by its integrated antivirus engines.

\vspace{0.02in}

\noindent\textbf{\circled{4} Identification of sensitive parameters in Docker run commands.} We analysed the execution instructions provided by component owners to identify \texttt{Docker run commands} containing sensitive parameters that could be exploited in security attacks. First, we examined serverless components hosted on Docker Hub for sensitive parameters previously documented in the literature~\cite{10.1007/978-3-030-58951-6_13}. Next, we extended our analysis and discovered three previously undocumented sensitive parameters that can pose significant security risks. These newly identified parameters could enable adversaries to escalate privileges, bypass security controls or gain unauthorized access, thus increasing the potential impact of compromised serverless components distributed via public repositories.

\vspace{0.02in}


\noindent\textbf{\circled{5} Finding misconfigurations in IaC templates.} We examined a large corpus of IaC templates to assess whether they contained insecure configurations that adversaries could exploit to compromise serverless applications. Using Trivy, we scanned these templates~\cite{trivy_iac} for misconfigurations across widely used frameworks, including Terraform, AWS CloudFormation and AWS SAM, and then classified the identified issues into four severity levels. For parameters frequently associated with security risks, we conducted an in-depth analysis to evaluate their potential impact and trace their underlying root causes. Additionally, we manually reviewed the IaC templates to discover previously undocumented misconfigurations that may not be detected through Trivy’s automated analysis.

\vspace{0.02in}

\noindent\textbf{\circled{6} Detection of potential typo-squatting attacks.} To identify potential typo-squatting attacks, we measured the similarity between component names using the Damerau-Levenshtein (DL) distance metric~\cite{277154}, which quantifies the minimum number of operations (insertions, deletions, substitutions, or transpositions) needed to transform one string into another. For each repository, we extracted both the username and image name associated with every serverless component and performed exhaustive pairwise comparisons to detect suspiciously similar naming patterns indicative of potential typo-squatting attacks. We focused on name pairs with low DL distances, as these indicate identical or highly similar names.





\begin{table*}[t]
  \centering
  \caption{Vulnerability statistics per repository using data collected from Trivy (\textcolor{blue}{blue}) and Grype (\textcolor{red}{red}), respectively.}
  \label{tab:mytable}
  \resizebox{\textwidth}{!}{
    \begin{tabular}{c c ccccc}
    \toprule
    \multirow{2}{*}{\textbf{Repository}} & \multirow{2}{*}{\textbf{\# of Compo.}} & \multicolumn{5}{c}{\textbf{\# of Vulnerabilities}}              \\ \cmidrule{3-7} 
    &                              & \textbf{Mean} & \textbf{Median} & \textbf{Max} & \textbf{Min} & \textbf{Std Dev} \\ \midrule
    Serverless Framework~\cite{Plugins}       & 355                     & \textcolor{blue}{8}/\textcolor{red}{35} & \textcolor{blue}{0}/\textcolor{red}{11} & \textcolor{blue}{257}/\textcolor{red}{611} & \textcolor{blue}{0}/\textcolor{red}{0} & \textcolor{blue}{24}/\textcolor{red}{70} \\
    AWS SAR~\cite{awsrepo}  & 242                     & \textcolor{blue}{4}/\textcolor{red}{21} & \textcolor{blue}{0}/\textcolor{red}{0} & \textcolor{blue}{127}/\textcolor{red}{3437} & \textcolor{blue}{0}/\textcolor{red}{0} & \textcolor{blue}{13}/\textcolor{red}{223} \\
    GitHub~\cite{git}                     & 712                     & \textcolor{blue}{98}/\textcolor{red}{119} & \textcolor{blue}{6}/\textcolor{red}{23} & \textcolor{blue}{18559}/\textcolor{red}{9658} & \textcolor{blue}{0}/\textcolor{red}{0} & \textcolor{blue}{838}/\textcolor{red}{581} \\
    Docker Hub~\cite{dockerhub}                 & 1374                    & \textcolor{blue}{1243}/\textcolor{red}{1522} & \textcolor{blue}{432}/\textcolor{red}{519} & \textcolor{blue}{6897}/\textcolor{red}{5781} & \textcolor{blue}{0}/\textcolor{red}{0} & \textcolor{blue}{1477}/\textcolor{red}{1628} \\
    Red Hat Quay~\cite{Red}               & 75                      & \textcolor{blue}{620}/\textcolor{red}{676} & \textcolor{blue}{135}/\textcolor{red}{230} & \textcolor{blue}{5390}/\textcolor{red}{5584} & \textcolor{blue}{0}/\textcolor{red}{0} & \textcolor{blue}{1057}/\textcolor{red}{1157} \\ \bottomrule
    \end{tabular}
    }
\end{table*}

\newpage

\section{Application-level Security Risks}

In this section, we assess the extent to which the collected serverless components are exposed to application-level security risks (RQ1). We begin by analysing the presence of known vulnerabilities in third-party libraries included in these components (\attackvector{V1}). We then investigate the potential for concealing malicious behaviour when these components are distributed in compressed formats (\attackvector{V2}).

\subsection{Vulnerability Analysis in Third-party Libraries}

\noindent\textbf{Statistical analysis of vulnerabilities across repositories.} To characterize the vulnerability landscape of each repository, we begin by reporting key statistical metrics---including the mean, median, minimum, maximum and standard deviation of vulnerability counts---based on data obtained from both Trivy and Grype (see Table~\ref{tab:mytable}). Our findings reveal a consistent discrepancy between both tools, with Grype systematically reporting higher vulnerability counts. This difference is rooted in their distinct design philosophies. Grype prioritizes sensitivity and broad detection coverage at the cost of a higher false positive rate~\cite{Chainguard},~\cite{Chainguard1}, while Trivy adopts a more conservative approach focused on minimizing false positives that may occasionally lead to missed vulnerabilities~\cite{trivyvulnerability}. Among the repositories we analysed, Docker Hub and Red Hat Quay exhibit the highest mean and median vulnerability counts, as well as the largest variability among components. GitHub falls in an intermediate position, with several components exhibiting significant vulnerabilities, though generally fewer than those in Docker Hub and Red Hat Quay. Conversely, we found that the Serverless Framework and AWS SAR consistently report lower vulnerability counts.

\vspace{0.02in}

\begin{figure}[t]
    \centering
    \begin{minipage}[t]{0.48\linewidth}
       \centering
       \includegraphics[width=\linewidth]{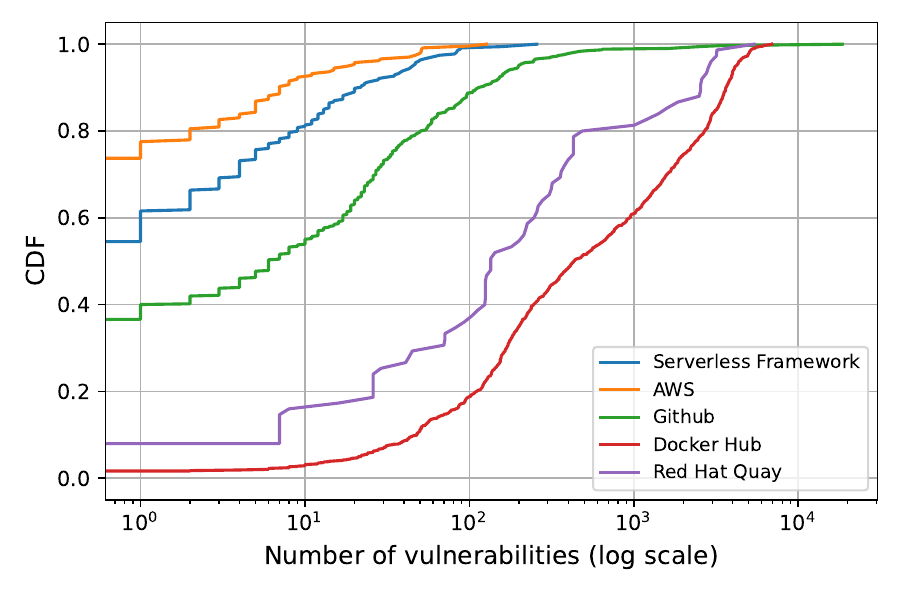}
        \caption{CDF of vulnerability counts per serverless component.}
        \label{fig:cdf}
    \end{minipage}
    \hfill
    \begin{minipage}[t]{0.48\linewidth}
        \centering
        \includegraphics[width=\linewidth]{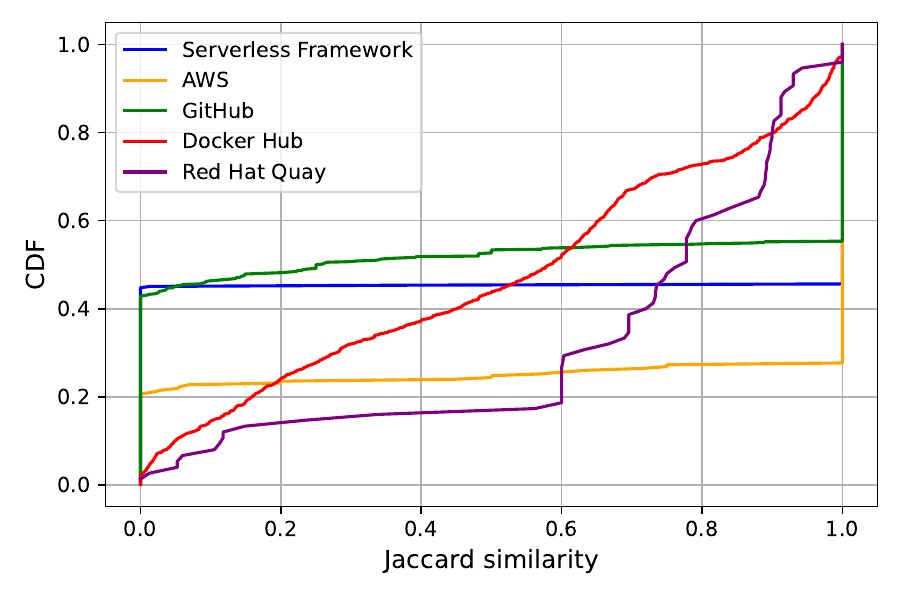}
        \caption{Trivy vs.\ Grype vulnerability similarity.}   \label{fig:Similarity_of_Vulnerabilities}
    \end{minipage}
    \vspace{-0.2in}
\end{figure}

\noindent\textbf{Distribution of vulnerabilities across serverless components.} To further understand the distribution of vulnerabilities within repositories, we analyse the Cumulative Distribution Function (CDF) of vulnerability counts per serverless component based on the data obtained from Trivy (see Figure~\ref{fig:cdf}). Our results show that approximately 80\% of Docker Hub components and 60\% of Red Hat Quay components have more than 100 vulnerabilities. GitHub presents a slightly better security posture: around 50\% of its components contain ten or fewer vulnerabilities, while 10\% have more than 100 vulnerabilities. By contrast, serverless components from the AWS SAR and Serverless Framework are significantly less affected. Many components have no known vulnerabilities and most affected ones contain fewer than 10, with few exceeding 100.



\noindent\textbf{Comparison of Grype and Trivy detection results.} Although Trivy and Grype use similar techniques for vulnerability detection, their results often differ significantly (see Table~\ref{tab:mytable}). To quantify this divergence, we computed the Jaccard similarity between the sets of vulnerabilities identified by each tool (see Figure~\ref{fig:Similarity_of_Vulnerabilities}). A score of 1 indicates complete agreement while a score of 0 indicates no overlap. For components from AWS SAR, GitHub and the Serverless Framework, many showed a Jaccard similarity of 1, suggesting identical results. Manual inspection revealed that these components frequently had no detected vulnerabilities. In contrast, components from Docker Hub and Red Hat Quay showed greater discrepancies, with fewer instances of high similarity. These findings underscore the importance of using multiple scanners to obtain a comprehensive assessment of security risks. Importantly, there is no universally accepted `ground truth' for vulnerability detection. Some organizations prioritize precision by considering only the intersection of scanner outputs to reduce false positives, while others prioritize recall by using the union of results to maximize coverage, even at the expense of increased false positives.

\begin{figure}[t]
\centering
\subfloat{
  \centering
  \includegraphics[width=.4\linewidth]{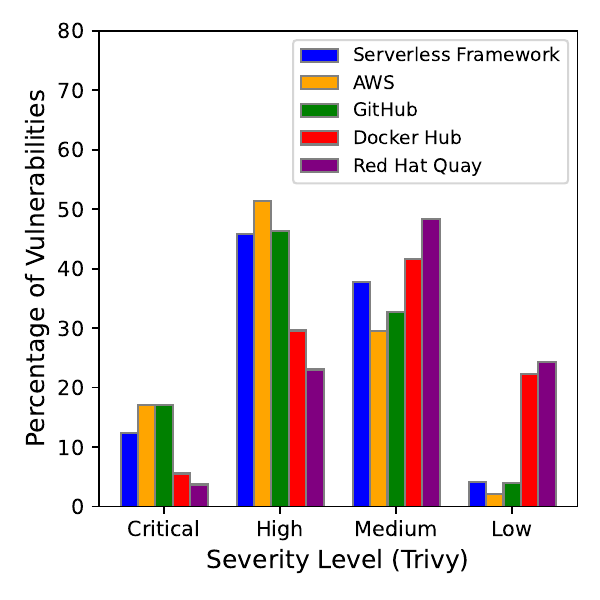}
}
\subfloat{
  \centering
  \includegraphics[width=.4\linewidth]{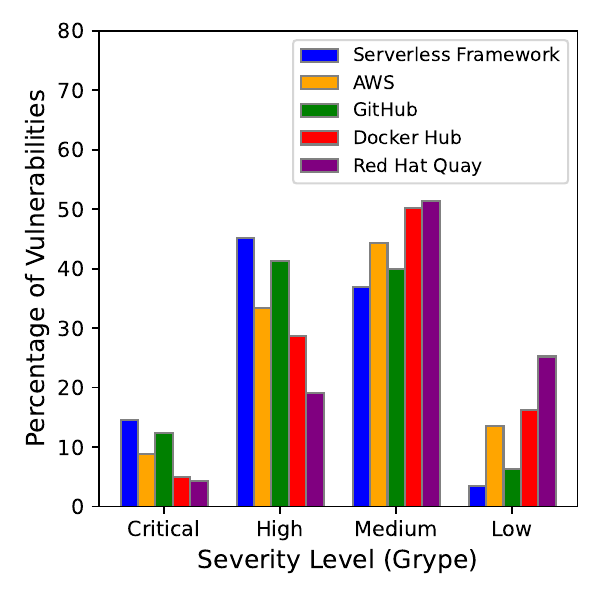}
}
  \caption{Vulnerability severity using data from Trivy (left) and Grype (right).}
  \label{fig:Severity}
  
\end{figure}

\noindent\textbf{Distribution of vulnerability severity across repositories.} While previous analyses focused primarily on vulnerability counts, we also examined vulnerability severity, a key factor in assessing real-world security risks (see Figure~\ref{fig:Severity}). Although Docker Hub and Red Hat Quay report the highest total vulnerabilities, the proportion of critical and high-severity vulnerabilities in their components is relatively low, only 5\% critical and 29\% high in Docker Hub, and 3\% critical and 23\% high in Red Hat Quay. In contrast, AWS SAR and Serverless Framework, which have the lowest average vulnerability counts per component, show the highest proportions of severe vulnerabilities: 68\% in AWS SAR and 57\% in Serverless Framework are classified as critical or high. For details on the top 10 most commonly used packages across five platforms and their associated vulnerabilities and severities, we refer the reader to Appendix~\ref{sec:appendix_vuln}.

\newcolumntype{x}{>{\centering\arraybackslash}m{0.2cm}}
\newcolumntype{h}{>{\centering\arraybackslash}m{1.5cm}}
\newcolumntype{i}{>{\centering\arraybackslash}m{1cm}}
\setlength{\tabcolsep}{4pt}


\begin{figure*}[t]
    \centering
    \subfloat[AWS SAR]{
        \includegraphics[width=0.3\textwidth]{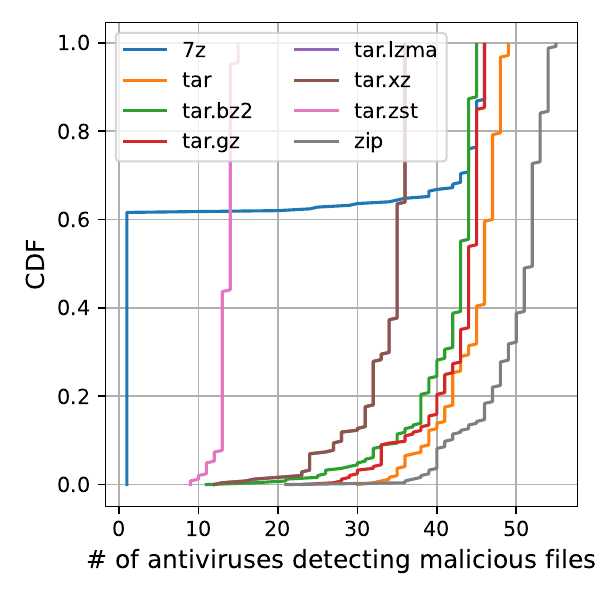}
    
    }
    \subfloat[GitHub]{
        \includegraphics[width=0.3\textwidth]{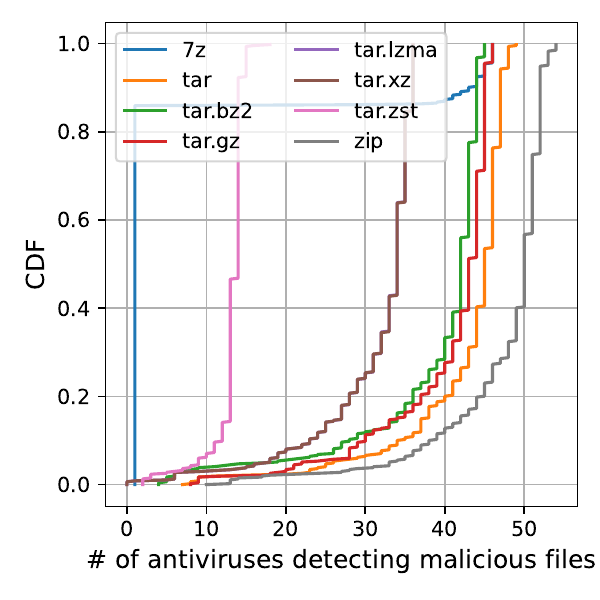}
    
    }
    \subfloat[Serverless Framework]{
        \includegraphics[width=0.3\textwidth]{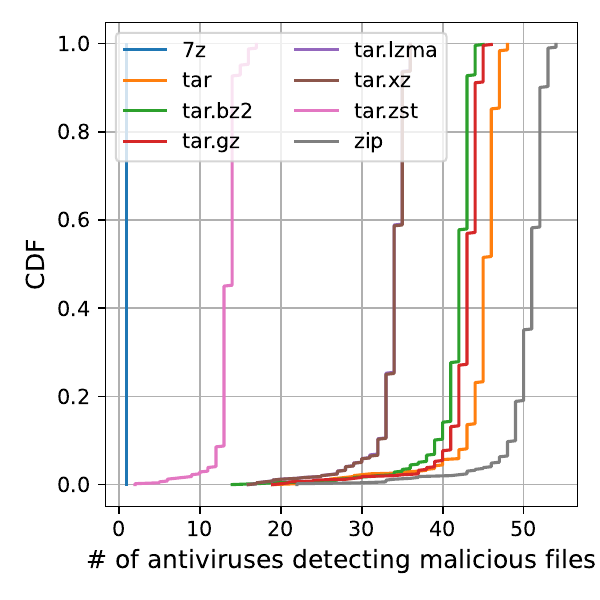}
    
    }
    
    \subfloat[Docker Hub]{
        \includegraphics[width=0.3\textwidth]{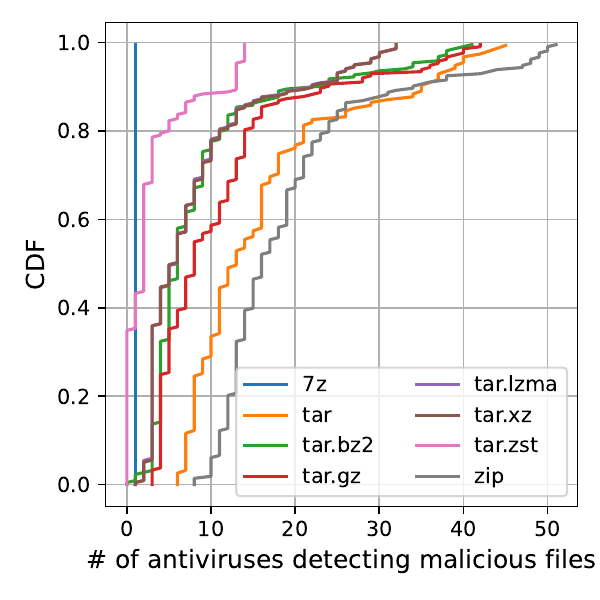}
    
    }
    \subfloat[Red Hat Quay]{
        \includegraphics[width=0.3\textwidth]{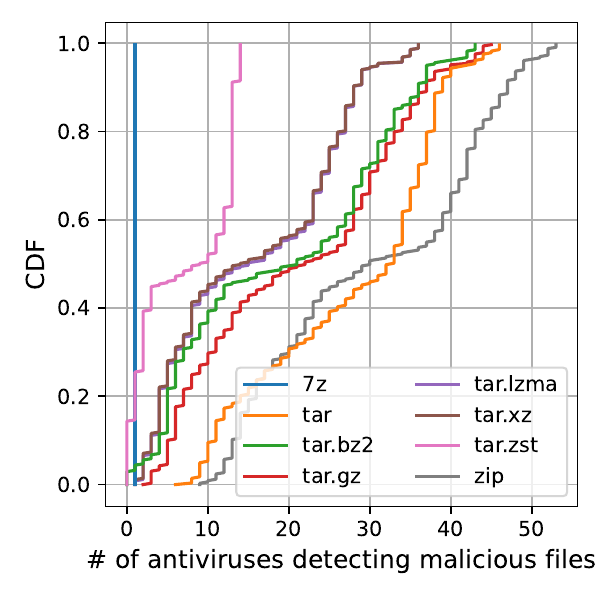}
    }
    
    \caption{Relationship between serverless components compressed with various compression algorithms and the number of engines that successfully detect them.}
    \label{fig:components_malicious_algorithms}
\end{figure*}

\subsection{Hiding Malicious Behaviour in Compressed Serverless Components}

To evaluate the potential for concealing malicious behaviour within compressed serverless components, we randomly selected 1,794 components from public repositories, including 356 from AWS SAR, 657 from GitHub, 352 from the Serverless Framework, 354 from Docker Hub and 75 from Red Hat Quay. Each component was compressed in its original form using eight widely adopted formats: \emph{7z}, \emph{tar}, \emph{tar.bz2}, \emph{tar.gz}, \emph{tar.lzma}, \emph{tar.xz}, \emph{tar.zst} and \emph{zip}. We first verified that none were flagged as malicious by VirusTotal. Then, using six malicious files sourced from reputable open-source malware repositories\footnote{For example, \url{https://github.com/vxunderground/MalwareSourceCode} and \url{https://github.com/nijithneo/DAT}}, including (i) the \texttt{eicar.txt} antivirus test file, (ii) a Python remote access trojan, (iii) a Java infector, (iv) a PHP backdoor, (v) a Python backdoor and (vi) a Python trojan, we generated malicious samples by injecting one file at a time into each component and recompressing them with each compression format.


We examined how the choice of compression format affects the detection of malicious behaviour. Figure~\ref{fig:components_malicious_algorithms} presents the CDF of antivirus engines that flagged malicious components for each format. Although all injected files were detected by at least one engine, components compressed with \emph{7z} and \emph{tar.zst} were flagged by significantly fewer engines, indicating lower detection reliability. This is concerning because, due to trade-offs between false positives and false negatives, a component is typically considered malicious only if flagged by a minimum number of engines. In prior work, this threshold was set at five engines~\cite{10.1007/978-3-030-58951-6_13}. In our analysis, we observed several instances where malicious components compressed with certain formats fell below this threshold, suggesting that embedding malicious behaviour within compressed serverless components could be an effective evasion technique. Additionally, we confirmed that such malicious components can be successfully uploaded to public serverless repositories (see Section~\ref{sec:discussion}).

\section{Configuration and Deployment Security Risks}

In this section, we examine the security risks in configuring and deploying serverless components (RQ2), focusing on three attack vectors: (i) sensitive parameters in Docker run commands (\attackvector{V3}); (ii) misconfigurations in IaC templates (\attackvector{V4}); and, (iii) potential typo-squatting attacks (\attackvector{V5}).

\subsection{Sensitive Parameter Misuse in Docker Run Commands}

We analysed the Docker run commands in the execution instructions for serverless components hosted on Docker Hub (\attackvector{V3}). We first targeted parameters previously identified as sensitive by Liu et al.~\cite{10.1007/978-3-030-58951-6_13}, including \texttt{-v}, \texttt{--privileged} and \texttt{--pid} which can compromise container isolation and host security. For instance, \texttt{-v <src>:<dest>} mounts host directories into a container, potentially leaking sensitive files; \texttt{--privileged} grants unrestricted access to host resources; and \texttt{--pid=host} enables container processes to monitor and interact with all host processes, facilitating reconnaissance or privilege escalation. Our analysis found 137 uses of \texttt{-v}, two instances of \texttt{--privileged} and no occurrences of \texttt{--pid}. \\

\noindent\textbf{Uncovering new risky parameters in Docker run commands.} Beyond assessing the prevalence of known sensitive parameters in Docker run commands for serverless components hosted in public repositories, our analysis uncovered three previously undocumented parameters that pose significant security risks. \\

\noindent\textit{(1) \underline{Hard-coded credentials.}} We discovered a case where AWS access keys were hard-coded directly into a Docker run command. Executing such commands may inadvertently expose sensitive credentials, allowing adversaries to take control of the associated cloud environment. This can lead to a range of attacks, including unauthorized access to S3 buckets or launching cryptojacking campaigns~\cite{elektra}. To mitigate this, we recommend avoiding credential injection via command-line arguments and instead using secure methods like Docker secrets (e.g., \texttt{--secret aws\_key} and \texttt{--secret aws\_secret}) to manage sensitive information.

\begin{lstlisting}
docker run -d --name go-serverless-aws-container   
-v $PWD:/usr/src/go/src 
-e <@\textcolor{myred}{AWS\_KEY=JFHGUFJAKEXAMPLEJDFJHEKF}@> 
-e <@\textcolor{myred}{AWS\_SECRET=AJDFUEXAMPLESDLKF}@>
-e AWS_REGION=us-east 
iamfrisbee/go-serverless-aws
\end{lstlisting}

\noindent\textit{(2) \underline{Sensitive information passed via environment variables.}} We found 47 instances where the \texttt{-e} parameter was used to pass sensitive information, such as credentials, tokens or keys, to containers. Adversaries with access to the Docker daemon (e.g., external adversaries who exploited a misconfiguration) can retrieve these values using commands like \texttt{docker inspect}. This exposure could lead to unauthorized access to cloud services, data exfiltration or financial abuse.
As with hard-coded credentials, this risk can be mitigated by using Docker secrets.
\begin{lstlisting}
docker run -v $(pwd):/opt/app 
<@\textcolor{black}{-e AWS\_DEFAULT\_REGION}@>
<@\textcolor{red}{-e AWS\_ACCESS\_KEY\_ID}@>
<@\textcolor{red}{-e AWS\_SECRET\_ACCESS\_KEY}@>
andrewoh531/docker-serverless serverless deploy
\end{lstlisting}

\noindent\textit{(3) \underline{Mounting the Docker daemon socket within containers.}} We identified a case where the Docker daemon socket (\texttt{/var/run/docker.sock}) was mounted directly into a container. This configuration effectively grants the container full control over the Docker daemon, allowing an adversary who compromises the container to escalate privileges and gain control over the host system. The security implications are comparable to those of the \texttt{--privileged} flag and are widely regarded as a critical misconfiguration.

\begin{lstlisting}
docker run -p 8080:8080 -v
<@\textcolor{red}{/var/run/docker.sock:/var/run/docker.sock}@>
furikuri/serverless-to-go
\end{lstlisting}

      

\begin{figure}[t]
    \centering
    \subfloat[AWS]{
        \includegraphics[width=.4\linewidth]{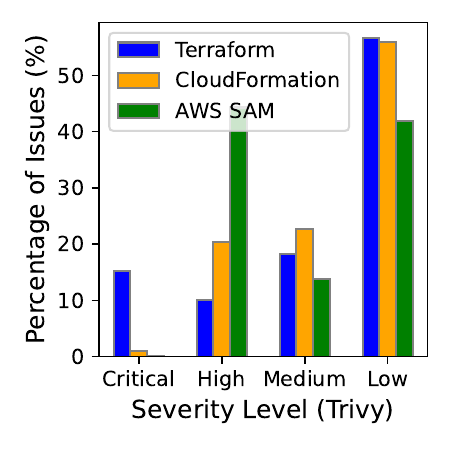}
    }
    \subfloat[GitHub]{
        \includegraphics[width=.4\linewidth]{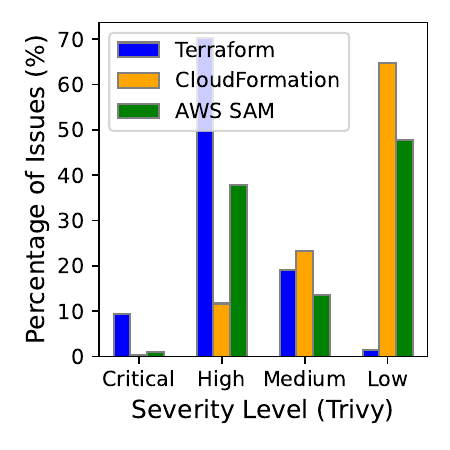}
    }
    \caption{Misconfigurations severity reported by Trivy for: (a) AWS Serverless Repository; and, (b) GitHub.}
    \label{fig:iac_severity}
\end{figure}

\subsection{Security Analysis of IaC Templates}

Given their critical role in automating the deployment of serverless applications, we conducted an in-depth analysis of IaC templates across three widely adopted frameworks: (i) Terraform; (ii) AWS CloudFormation; and, (iii) AWS Serverless Application Model (SAM) (\attackvector{V4}). Our dataset includes IaC templates from AWS SAR and GitHub, as container-based platforms such as Docker Hub, Red Hat Quay and Serverless Framework do not include IaC templates. Table~\ref{tab:iac_stats} provides a breakdown of the analysed IaC templates. To distinguish between frameworks, we classified templates by their extensions: \texttt{.tf} for Terraform and \texttt{.json} or \texttt{.yaml} for AWS CloudFormation. AWS SAM templates, which also use the \texttt{.yaml} format, were identified by detecting the presence of the \textit{Transform: AWS::Serverless-2016-10-31} directive.

\begin{table}[t]
    \centering
    \scriptsize
    \caption{Breakdown of IaC templates analyzed in our serverless repository dataset.}
    \begin{tabular}{c c c c c}
    
        \toprule
         \textbf{Repository} & \makecell{\textbf{Terraform}\\(\texttt{.tf})} & \makecell{\textbf{CloudFormation}\\(\texttt{.yaml}, \texttt{.json})} & \makecell{\textbf{AWS SAM}\\(\texttt{.yaml})} & \textbf{Total} \\ \midrule
         
         AWS & 30 & 5,023 & 225 & 5,278 \\ \midrule
         GitHub & 1,764 & 117,860 & 1,034 & 120,658 \\ \midrule
         \textbf{Total} & 1,794 & 122,883 & 1,259 & 125,936 \\ \bottomrule
    \end{tabular}
    \vspace{-0.2in}
    \label{tab:iac_stats}
\end{table}

We scanned the retrieved IaC templates using Trivy to identify potential misconfigurations. Figure~\ref{fig:iac_severity} shows the distribution of misconfiguration issues across four severity levels. Our analysis shows that Terraform IaC templates exhibit a significantly higher proportion of critical misconfigurations than AWS CloudFormation and AWS SAM templates. This disparity is likely due to Terraform’s broader configurability across multi-cloud environments. Nevertheless, across all frameworks, the proportion of critical and high-severity issues remains substantial, underscoring the systemic risk posed by IaC misconfigurations. We also examined the five most common misconfigurations found in the templates. The most frequent issue, accounting for approximately 19\% of the identified misconfigurations, involved the omission of a source ARN in Lambda permissions~\cite{ensure_terraform}, potentially allowing unrestricted invocation of Lambda functions. Another common issue, accounting for roughly 13\% of cases, was the use of default AWS-managed keys instead of customer-managed encryption keys for S3 buckets~\cite{managed_keys}, which weakens control over data protection. A complete overview of the identified misconfigurations is given in Appendix~\ref{sec:appendix_iac}. \\


\noindent\textbf{Cross-Origin Resource Sharing}. It is well known that using a wildcard (`*') in Cross-Origin Resource Sharing (CORS) policies introduces security risks~\cite{chen2018we}. Our analysis revealed such misconfigurations in IaC templates configuring CORS for both AWS CloudFormation and AWS SAM. Through manual inspection, we identified seven instances of this issue in AWS SAR and one in GitHub, which were \emph{not} detected by Trivy. The \texttt{CorsOrigin} attribute is used to configure CORS policies, which control cross-domain resource sharing (e.g., allowing front-end applications to access resources from AWS S3 buckets)~\cite{cors}. However, we found that several templates included the configuration \texttt{Default: `*'}, which allows unrestricted access from any domain to the serverless application’s API Gateway. This wildcard disables CORS protections, posing a significant risk. If adversaries compromise a serverless function, they can exploit the permissive CORS policy to exfiltrate sensitive data to untrusted or attacker-controlled domains~\cite{datta2022alastor}. While browsers block credentialed requests (e.g., those involving cookies or authorization headers)~\cite{browser_cred}, our manual inspection of these IaC templates revealed that they provision public APIs without authentication, thereby making even unauthenticated cross-origin requests a security risk.


\subsection{Typo-squatting Attacks}

As the final step in our analysis, we assessed the susceptibility of public serverless repositories to typo-squatting attacks (\attackvector{V5}), covering components from the Serverless Framework, GitHub, Red Hat Quay and Docker Hub. AWS SAR was excluded since its components are used exclusively within AWS, where the chance of typos by developers is much lower.

\begin{figure}[t!]
\centering
    \subfloat{
      \centering
      \includegraphics[width=0.4\linewidth]{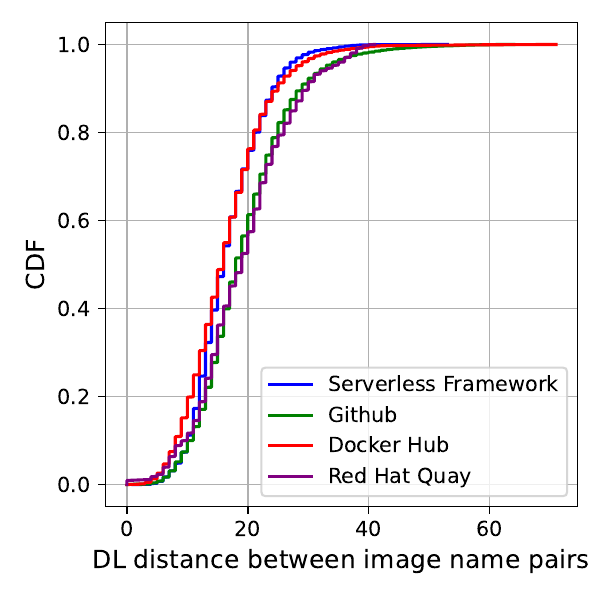}
      \label{fig:lex1}
    } 
    \subfloat{
      \centering
      \includegraphics[width=0.4\linewidth]{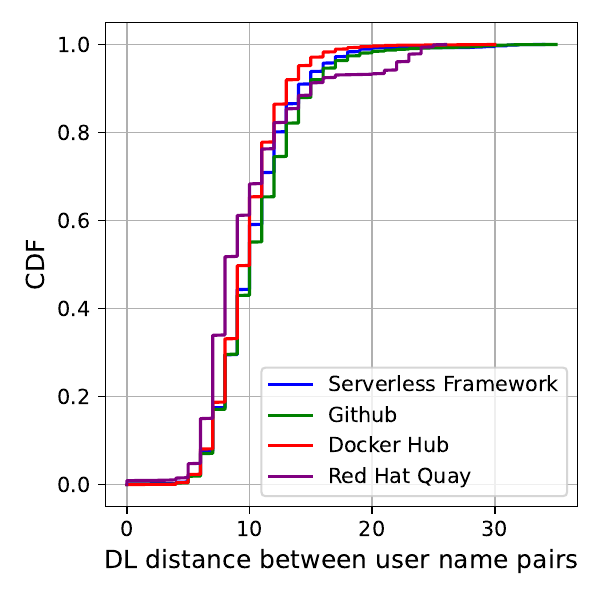}
      
      \label{fig:lex2}
    }
    \captionof{figure}{Lexical analysis based on Damerau-Levenshtein distance, applied to image names (left) and user names (right)}
    \label{fig:lex}
    \vspace{-0.2in}
\end{figure}

Figure~\ref{fig:lex} shows the CDF of DL similarities for usernames and image names. Our findings revealed two pairs of \emph{user names} with a DL distance equal to 1: one identified on GitHub and one on Docker Hub. Additionally, we identified several \emph{image name} pairs with a DL distance of 1, including two instances in the Serverless Framework, six in GitHub, 191 in Docker Hub and one in Red Hat Quay. These results indicate that while typo-squatting risks are present across all repositories, they are particularly pronounced in Docker Hub. Although the intent behind similar usernames or image names within the same repository is unclear, their presence suggests this vector could be exploited by adversaries.

\newpage

\section{Key Findings and Discussion}
\label{sec:discussion}

Our analysis reveals that public serverless repositories often lack rigorous security oversight. Repository administrators typically perform minimal (if any) security checks on the components they host. This is particularly concerning in serverless computing, where applications are rapidly built from many small, interconnected serverless functions. While this enables fine-grained security controls when properly implemented, it also requires each function to be independently configured and secured, increasing the risk of errors and misconfigurations.

\vspace{0.05in}

\noindent\textbf{Application-level security risks.} Although vulnerabilities in third-party libraries are a well-known threat, our findings show that such issues remain widespread even in repositories presumed to maintain strict security controls like AWS SAR. Beyond these, we uncovered new risks tied to how serverless applications are deployed. Specifically, the ability to upload pre-packaged ZIP archives bundling code and dependencies creates a blind spot for traditional scanners, which often fail to properly inspect compressed artifacts. To demonstrate this, we uploaded a malicious ZIP archive with known malware to AWS SAR via a controlled test repository. The upload completed without triggering any alerts, exposing a significant gap in current threat detection. While we deleted the component immediately to comply with ethical research standards, this experiment highlights the need for more advanced security techniques to detect malicious content within compressed serverless packages. 

\vspace{0.05in}

\noindent\textbf{Configuration and deployment security risks.} Our study also reveals significant risks arising from serverless application configuration and deployment practices. A common issue is the widespread use of default, overly permissive, or publicly recommended settings. Specifically, both Docker run commands and IaC templates frequently embed risky parameters, making them untrustworthy by default. Although these configuration-related threats are somewhat fewer than third-party library vulnerabilities, they are much harder to detect, as shown by our discovery of three new sensitive Docker parameters and a novel IaC misconfiguration that evaded detection by advanced scanners like Trivy. Furthermore, serverless repositories are vulnerable to typo-squatting attacks, where adversaries register components with names closely resembling popular ones. Critically, serverless automation amplifies the impact of even rare typos, since IaC templates and CI/CD pipelines may automatically fetch and deploy typo-squatted components without manual review, enabling silent integration into production.

\vspace{0.02in}

\noindent\textbf{Security model.} We advocate for a security model in which public serverless repository administrators are the primary responsible for securing the components they host. To ensure transparency and accountability, they should disclose their security practices, including scanning methods, tools and the frequency of their security assessments.  Administrators must implement automated, periodic vulnerability scans using a diverse set of complementary tools, alongside mandatory pre-publication compliance checks. Given the rapidly evolving serverless ecosystem, with frequent emergence of new vulnerabilities in libraries, frameworks and cloud services, continuous monitoring and automated re-evaluation of hosted components are crucial for sustained security. Additionally, repository maintainers should apply automated static analysis to configuration files and enforce blacklists of sensitive parameters (such as those identified in this and prior work) to prevent their inclusion. To mitigate typo-squatting attacks, mechanisms like similarity-based name collision alerts and stricter naming policies should be adopted. Until these security measures become standard, users and developers should assume that any serverless component from a public repository may be compromised and perform thorough independent security evaluations before use.

\vspace{0.02in}

\noindent\textbf{Possible limitations of this work.} False positives are a well-known limitation of vulnerability scanners such as Trivy and Grype. These tools may classify a repository as vulnerable if it includes a package with known CVEs in its metadata files. However, if the package is not actually used in the source code, the reported vulnerability may constitute a false positive. To assess the extent of this issue, we randomly selected 30 open-source serverless components (approximately the square root of the total: 242 from AWS and 712 from GitHub). Trivy and Grype collectively reported 1,417 CVEs across these samples. We then manually verified whether the flagged packages were actually referenced in the source code. If a package appeared only in metadata files (e.g., \texttt{.lock}, \texttt{.gradle}, \texttt{.toml}, \texttt{.yml}, \texttt{.yaml}, \texttt{.json}, \texttt{.xml}, \texttt{.md}) but not in source code files (e.g., \texttt{.py}, \texttt{.js}, \texttt{.ts}, \texttt{.java}, \texttt{.go}, \texttt{.rb}, \texttt{.sh}, \texttt{.c}, \texttt{.cpp}), we classified the associated CVE as a false positive. Our analysis found that 1,275 out of 1,417 CVEs were associated with packages present in the source code, resulting in an estimated false positive rate of approximately 10\%. These findings suggest that while false positives are present, the rate is within acceptable bounds and does not undermine the overall reliability of our vulnerability analysis.

\section{Related work}



Shu \emph{et al.}~\cite{10.1145/3029806.3029832} were the first to examine the security state of Docker Hub images. Wist \emph{et al.}~\cite{wist2020vulnerability} conducted a similar study on 2,500 Docker Hub images. Liu \emph{et al.}~\cite{10.1007/978-3-030-58951-6_13} provided a comprehensive assessment of the Docker Hub ecosystem, focusing on the detection of malicious images and the identification of exploitable parameters in Docker run commands. Other researchers have focused on analysing specific types of Docker Hub images. For instance, Zerouali \emph{et al.}~\cite{8668013,10.1007/s10664-020-09908-6} analysed vulnerabilities and outdated packages in Debian-based and programming language-specific images~\cite{ZEROUALI2021102653} and Haque \emph{et al.}~\cite{DBLP:journals/corr/abs-2112-12597} evaluated the exploitability of vulnerabilities in base images.

\vspace{0.02in}

\noindent\textbf{Research gap.} Prior work has mainly addressed security issues in microservices distributed via Docker Hub. In contrast, our study focuses on serverless computing, which is rapidly becoming the dominant cloud application deployment model. We evaluated the vulnerability of serverless components to five distinct attack vectors, including two newly identified in this study, using components from five prominent public repositories. By analysing multiple repositories and diverse attack vectors, we provide a representative overview of current security practices in public serverless repositories.


\section{Conclusion}

This paper presents the first large-scale empirical analysis of security risks in public serverless repositories. Our study reveals systemic weaknesses across these repositories including (i) widespread use of vulnerable third-party dependencies; (ii) misconfigurations in IaC templates; (iii) sensitive parameters in Docker run commands; (iv) the ability to conceal malicious payloads in compressed serverless components; and (v) exposure to typo-squatting attacks. Based on these findings, we offer actionable recommendations for repository maintainers, developers and users to enhance the security of public serverless repositories.

\vspace{0.02in}

\noindent\textbf{Acknowledgments}. We thank the anonymous reviewers for their insightful feedback and help in improving this paper. This research received funding from the Smart Networks and Services Joint Undertaking (SNS JU) under the European Union’s Horizon Europe programme: ELASTIC (GA\#101139067); Horizon Europe: FLUIDOS (GA\#101070473) and LAZARUS (GA\#101070303); and the UNICO I+D Cloud program funded by the Ministry of Economic Affairs and Digital Transformation and the European Union–NextGenerationEU within the framework of the Plan de Recuperación, Transformación y Resiliencia (PRTR) with the CLOUDLESS project. This work was partially supported by project SERICS (PE00000014) under the NRRP MUR program funded by the EU - NGEU. Additionally, this work was partly supported by the National Research Foundation of Korea (NRF) grant funded by the Korea government (MSIT) (No. RS-2024-00457937, Design and implementation of security layers for secure WebAssembly-based serverless environments). 

\newpage

\bibliographystyle{splncs04}
\bibliography{references}

\appendix

\section{Vulnerability Composition}
\label{sec:appendix_vuln}

\newcolumntype{x}{>{\centering\arraybackslash}m{0.2cm}}
\newcolumntype{h}{>{\centering\arraybackslash}m{1cm}}
\newcolumntype{i}{>{\centering\arraybackslash}m{1cm}}
\setlength{\tabcolsep}{4pt}  

\begin{table*}[h]
\centering
\caption{Top 10 most commonly used packages across five platforms, along with the number and severity of associated vulnerabilities (C: \emph{Critical}, H: \emph{High}, M: \emph{Medium}, L: \emph{Low}).}
{\scriptsize
\resizebox{\textwidth}{!}{
\begin{tabular}{h i x x x x h i x x x x h i x x x x h i x x x x h i x x x x h i x x x x}

\toprule
\multicolumn{6}{c}{\textbf{AWS Serverless Repository}} & \multicolumn{6}{c}{\textbf{GitHub}} & \multicolumn{6}{c}{\textbf{Serverless Framework}} & \multicolumn{6}{c}{\textbf{Docker Hub}} & \multicolumn{6}{c}{\textbf{Red Hat Quay}} \\ \cmidrule(lr){1-6} \cmidrule(lr){7-12} \cmidrule(lr){13-18} \cmidrule(lr){19-24} \cmidrule(lr){25-30} \cmidrule(lr){31-36}

\textbf{Package} & \textbf{Rate} & \textbf{C} & \textbf{H} & \textbf{M} & \textbf{L} & \textbf{Package} & \textbf{Rate} & \textbf{C} & \textbf{H} & \textbf{M} & \textbf{L} & \textbf{Package} & \textbf{Rate} & \textbf{C} & \textbf{H} & \textbf{M} & \textbf{L} & \textbf{Package} & \textbf{Rate} & \textbf{C} & \textbf{H} & \textbf{M} & \textbf{L} & \textbf{Package} & \textbf{Rate} & \textbf{C} & \textbf{H} & \textbf{M} & \textbf{L} \\
\midrule

xml2js & 4.96\% & 0 & 0 & 1 & 0 & semver & 27.53\% & 0 & 0 & 2 & 0 & xml2js & 19.44\% & 0 & 0 & 1 & 0 & tar & 74.09\% & 0 & 7 & 6 & 2 & ncurses-base & 82.67\% & 0 & 1 & 5 & 21 \\ 
follow-redirects & 4.55\% & 0 & 1 & 2 & 0 & qs & 23.74\% & 0 & 4 & 1 & 0 & lodash & 19.15\% & 1 & 4 & 2 & 1 & semver & 63.97\% & 0 & 0 & 2 & 0 & ncurses-libs & 74.67\% & 0 & 0 & 5 & 22 \\
axios & 4.13\% & 0 & 2 & 2 & 0 & follow-redirects & 21.63\% & 0 & 1 & 2 & 0 & semver & 18.03\% & 0 & 0 & 1 & 0 & curl & 61.72\% & 25 & 34 & 47 & 16 & libgcc & 68.00\% & 0 & 0 & 12 & 14 \\
lodash & 3.72\% & 1 & 3 & 1 & 0 & xml2js & 21.21\% & 0 & 0 & 1 & 0 & minimatch & 17.46\% & 0 & 3 & 0 & 0 & minimatch & 57.50\% & 0 & 3 & 0 & 0 & pcre2 & 62.67\% & 0 & 0 & 2 & 1 \\
minimist & 3.72\% & 1 & 0 & 1 & 0 & minimist & 21.07\% & 1 & 0 & 1 & 0 & qs & 13.80\% & 0 & 1 & 0 & 0 & ncurses-base & 57.35\% & 2 & 6 & 13 & 6 & curl & 56.00\% & 13 & 10 & 46 & 24 \\
urllib3 & 3.72\% & 0 & 3 & 5 & 0 & minimatch & 20.22\% & 0 & 3 & 0 & 0 & minimist & 13.52\% & 1 & 0 & 1 & 0 & got & 57.21\% & 0 & 0 & 1 & 0 & ca-certificates & 54.67\% & 0 & 2 & 0 & 1 \\
minimatch & 2.89\% & 0 & 1 & 0 & 0 & lodash & 19.52\% & 1 & 4 & 2 & 1 & aws-sdk & 11.83\% & 0 & 1 & 0 & 0 & qs & 57.21\% & 0 & 4 & 1 & 0 & libxml2 & 52.00\% & 2 & 1 & 31 & 7 \\
aws-sdk & 2.89\% & 0 & 1 & 0 & 0 & tough-cookie & 19.10\% & 0 & 1 & 1 & 0 & axios & 9.01\% & 0 & 2 & 2 & 0 & openssl & 56.77\% & 4 & 29 & 84 & 34 & tar & 52.00\% & 0 & 7 & 3 & 5 \\
qs & 2.48\% & 0 & 1 & 0 & 0 & axios & 18.96\% & 0 & 2 & 2 & 0 & follow-redirects & 9.01\% & 0 & 1 & 2 & 0 & minimist & 54.29\% & 1 & 0 & 1 & 0 & glib2 & 48.00\% & 0 & 2 & 48 & 20 \\
semver & 2.48\% & 0 & 0 & 1 & 0 & node-fetch & 15.73\% & 0 & 1 & 1 & 1 & async & 8.45\% & 0 & 1 & 0 & 0 & xml2js & 53.28\% & 0 & 0 & 1 & 0 & gnupg2 & 48.00\% & 0 & 1 & 4 & 6 \\
\bottomrule
\end{tabular}
}
\label{tab:top_10}
}
\end{table*}

To gain deeper insight into the vulnerability landscape, we analysed the top 10 most commonly used packages in each repository (see Table~\ref{tab:top_10}). Notably, the \emph{lodash} and \emph{minimist} packages, each affected by one critical vulnerability (\texttt{CVE-2019-10744} and \texttt{CVE-2021-44906}, respectively) appear in three and four of the analysed repositories, respectively. A cross-repository comparison reveals that versions of these packages hosted in the AWS SAW consistently exhibit fewer vulnerabilities, suggesting that AWS may actively patch or curate its hosted packages. In contrast, Docker Hub and Red Hat Quay show significantly higher vulnerability counts for the same packages, likely due to frequent image reuse and less stringent update practices. Among all analysed packages, \emph{curl} stands out for both its widespread use—particularly in Docker Hub and Red Hat Quay—and its high number of critical and high-severity vulnerabilities.


\section{Common Misconfigurations in IaC Templates}
\label{sec:appendix_iac}

\begin{figure}[!h]
    \centering
    \vspace{-0.2in}
    \subfloat[Terraform in AWS]{
        \includegraphics[width=0.3\linewidth]{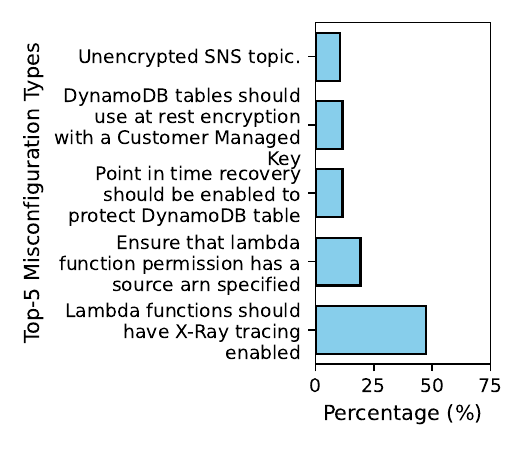}
    }
    \subfloat[Terraform in GitHub]{
        \includegraphics[width=0.3\linewidth]{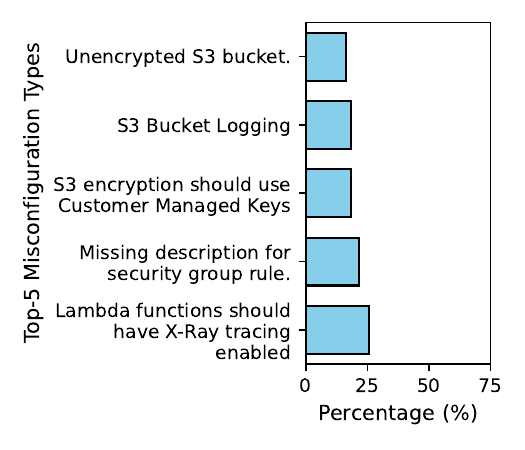}
    }
    \subfloat[CloudFormation in AWS]{
        \includegraphics[width=0.3\linewidth]{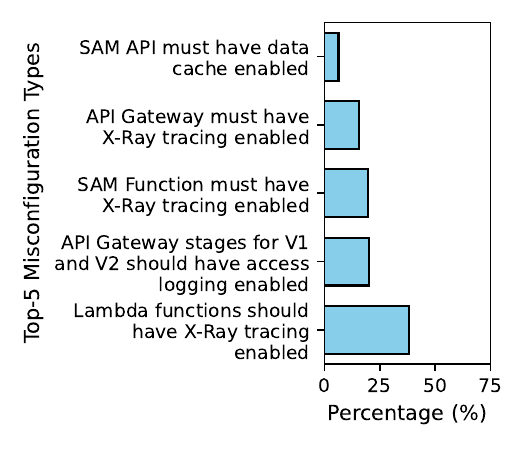}
    }
    
    \subfloat[CloudFormation in GitHub]{
        \includegraphics[width=0.3\linewidth]{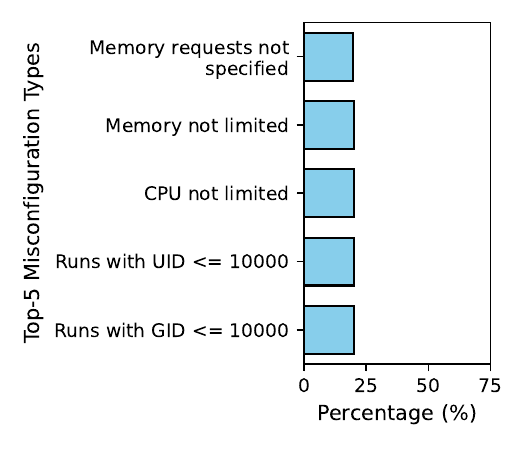}
    }
    \subfloat[SAM in AWS]{
        \includegraphics[width=0.3\linewidth]{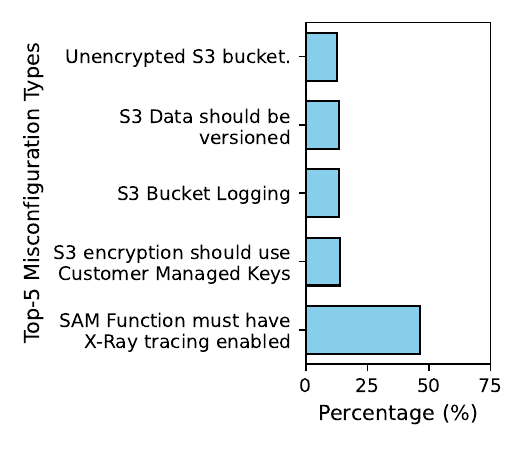}
    }
    \subfloat[SAM in GitHub]{
        \includegraphics[width=0.3\linewidth]{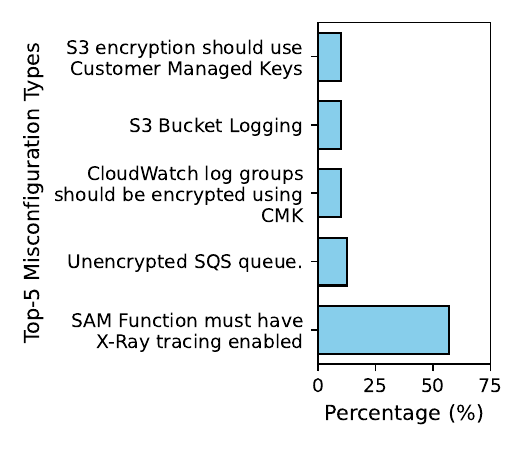}
    }
    
    \caption{Top-5 most common misconfigurations in IaC templates from AWS SAR and GitHub.}
    \label{fig:iac_popularity}
    \vspace{-0.2in}
\end{figure}

Figure~\ref{fig:iac_popularity} presents the five most frequently observed misconfigurations across the three analyzed IaC frameworks: Terraform, AWS CloudFormation, and AWS Serverless Application Model (SAM). The obtained results highlight recurring security issues that affect the security posture of serverless deployments.

\end{document}